\documentclass{llncs}
\usepackage{makeidx}  
\usepackage{graphicx}
\usepackage{here} 
\usepackage{amsmath}

\title{Simulation of mitochondrial metabolism using multi-agents system}

\author{Charles Lales\inst{1}$^{\&}$\inst{2} \and Nicolas Parisey\inst{2} \and Jean-Pierre
Mazat\inst{2} \and Marie Beurton-Aimar\inst{1}$^{\&}$\inst{2}}

\institute{ Univ. Bordeaux 1, LaBRI CNRS UMR 5800, France \and Univ.
Bordeaux 2, Mitochondrial Physiopathology Lab. INSERM U688, France\\
\email{lales@labri.fr}\\
\email{nicolas.parisey@etud.u-bordeaux2.fr}\\
\email{jpm@u-bordeaux2.fr}\\
\email{aimar@labri.fr}}

\begin{document}

\maketitle              

\begin{abstract}
Metabolic pathways describe chains of enzymatic reactions. Their modelling is a
key point to understand living systems. An enzymatic reaction is an interaction
between one or several metabolites (substrates) and an enzyme (simple protein
or enzymatic complex build of several subunits). In our Mitochondria in Silico
Project, MitoScop, we study the metabolism of the mitochondria, an
intra-cellular organelle.

Many ordinary differential equation models are available in the literature.
They well fit experimental results on flux values inside the metabolic
pathways, but many parameters are difficult to transcribe with such models:
localization of enzymes, rules about the reactions scheduler, etc Moreover, a
model of a significant part of mitochondrial metabolism could become very
complex and contain more than 50 equations.

In this context, the multi-agents systems appear as an alternative to model the
metabolic pathways. Firstly, we have looked after membrane design. The
mitochondria is a particular case because the inner mitochondrial space, ie
matricial space, is delimited by two membranes: the inner and the outer one. In
addition to matricial enzymes, other enzymes are located inside the membranes
or in the inter-membrane space. Analysis of mitochondrial metabolism must take
into account this kind of architecture.
\end{abstract}

\section{Introduction}
Metabolic pathways are a set of reactions catalyzed by enzymes which takes
place inside cells. They constitute the cellular metabolism. In biology,
numerical simulations allow interactions between molecules to test hypotheses
about normal or pathologic behaviours. Metabolism can be viewed as a network.
We have studied mitochondria, a specific intra-cellular organelle, which has
its own enzymes and thus its own network, generally analyzed independently of
the cell metabolism.

A classical way of modeling reaction chains is to use ordinary differential
equations (ODE). With such models it is possible to describe the evolution of
the system. But many biological aspects are difficult to describe with ODE:
particular situation where there are few molecules reacting (no average
behaviour), influence of the mass or the conformation of the molecules,
transient association of several molecules. However such examples correspond to
biological cases like disturbance of the network caused by lack of some
molecules during a small portion of time or genetic mutations which change
enzyme conformation. All of this shows that new models for simulation like
these using \textbf{multi-agents systems} may improve the understanding of
biological processes.
\paragraph{}
One way to design a biological network is to see it as a world filled with more
or less complex entities. One entity could be a simple molecule like a
metabolite or an enzymatic complex. Each reaction is described by interactions
between entities belonging to this world. In our model, each entity is
represented by a reactive agent. This agent is situated in a 3D space and the
reactions are scheduled by time steps.
\paragraph{}

In this paper, we show what is needed to model metabolism pathways with a
multi-agents system (MAS) in the context of the mitochondrial metabolism. We
present a first approach to a 3D molecular design and the application to the
particular case of the two mitochondrial membranes: the inner and the outer
one. Each of them is composed of a phospholipids bilayer containing enzymatic
complexes. The attention is focused on a generic description of molecular
interactions in order to use this behaviour both for simple case as lipids and
for more complex one like respiratory chain enzymes.

\section{Mitochondria: their membranes and their energetic metabolism}
\begin{figure}[H]
    \begin{center}
    \includegraphics[scale=0.7]{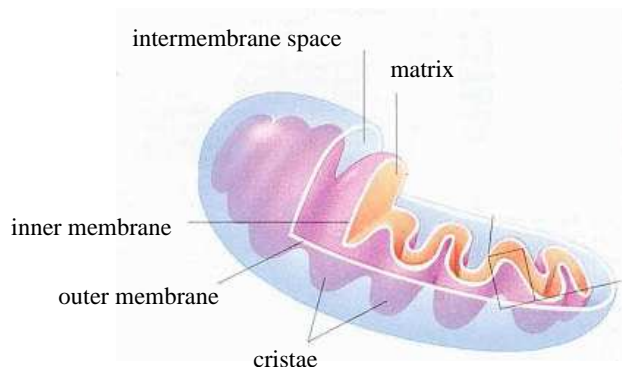}
    \end{center}
    \caption{\label{mito}Mitochondria from http://universe-review.ca/}
\end{figure}

Mitochondria (see Fig.\ref{mito}) are intra-cellular organelles which are the
power plant of the cell. They are delimited by a double membrane, composed of
phospholipids, which are naturally arranged in a bilayer to minimize their
interactions with water. The energy delivered by mitochondria comes from a
complex mechanism, involving a series of oxido-reduction reactions called
\textit{"Respiratory chain"} using oxygen we breathe in and the nutriments we
ingest. They are catalyzed by four highly structured protein complexes and
electron transporters (CoenzymeQ and CytochromeC) embedded in the inner
mitochondrial membrane. Electrons transfer is linked to an extrusion of protons
outside the mitochondria, creating an electro-chemical gradient, which will be
used for ATP synthesis and other works \cite{Maz05}\cite{Mit61}. This process
involves the respiratory chain and ATP synthesis, both of them called
"Oxidative phosphorylation". Oxidative phosphorylation has been modelled for a
long time in order to integrate all the kinetic and thermodynamic properties of
chemiosmosis theory developed by Peter Mitchell \cite{Mit61}.

However some aspects, particularly the electron transports between complexes I
or II and complex III of the respiratory chain by a molecule called CoQ
presenting different redox states, is rather complex \cite{Mit76}. CoQ is
certainly a central point in electron transport in respiratory chain. This
transport appears rather difficult to model with ODE \cite{Dem98} due to the
great variety of interactions, of the reactions involved at different reacting
sites and of the amount of the various redox CoQ species. In addition, these
reactions take place inside the inner mitochondrial membrane which is a highly
heterogeneous medium (see Fig.\ref{mito}).

\paragraph{}
All these questions could be addressed  with a multi-agents model. In fact
these problems illustrate how an MAS approach is highly suited to take care of
spatial disposition of the various enzymatic complexes, competition between
different complexes to access to some finite resources, complex interactions
between several types of molecules (membranes and enzymes) sometimes in low
quantity, environment effects (locally or not).

In the next section, we firstly describe the generic characteristics of the
retained MAS. Finally, we describe the specific molecular modeling (section 4).

\section{A multi-agents model for metabolic pathways}

Our MAS uses a 3D continuous space  where the agents are situated.
\textit{Situated} Agents use space as modality of their interactions. The
distances between agents are used to describe neighbouring rules. All the
agents have a local perception of their neighbourhood. A grid divides space
into neighbouring units which simplify interactions complexity giving directly
neighbour positions.

The MAS is called \textit{open} because of the variable number of the
agent during simulation. The continuous degradation and synthesis of molecules is a common biological process.

The time is discretized and a scheduler is defined for each type of agents.
Actually, the inner-mitochondrial membrane is composed of enzymes and
phospholipids. An enzymatic reaction should not be simulated on the same
time-scale as two phospholipids interaction. One can also notice that it is
possible to simulate time continuity with a discretization based on events and
not on time steps.

\begin{figure}[H]
  \begin{center}
    \includegraphics[scale=0.6]{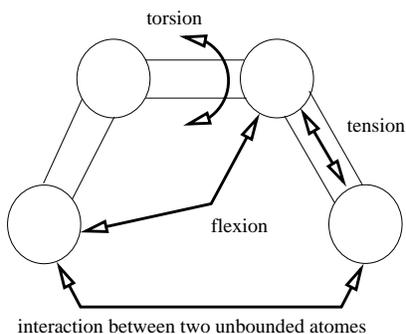}
    \caption{\label{force}Potentials deformations applied to a simple molecule.}
  \end{center}
\end{figure}

In general, the level used for representing the biological objects, i.e. the
model granularity, varies from atom to molecular groups (see \cite{Hut04} for a
short review). One can notice that an abstraction by a single point is not
sufficient to take into account a dynamic 3D structure and its spacial
orientation (see Fig.\ref{force}). On the contrary, molecular modeling
simulation (atom by atom) is not possible for a whole organelle due to the
large number of entities to compute.

\begin{figure}[H]
    \begin{center}
        \includegraphics[scale=0.6]{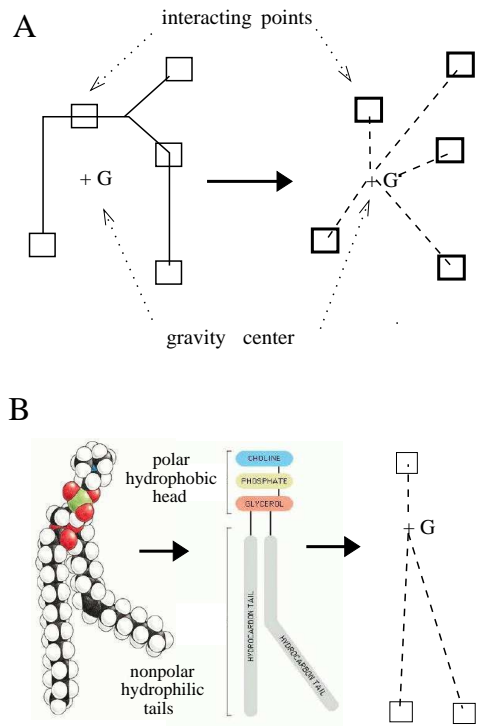}
        \caption{\label{phospho}A: abstracted 3D structure. B: application to a phospholipid.}
    \end{center}
\end{figure}

A phospholipids model published by Edwards \cite{Ed98} uses two interacting
points per molecule. In order to design a more generic model, we have extended
this previous model.  Our model is suitable for modeling metabolites and
enzymes alike. Thus we have chosen to reduce each agent (molecule) to its
gravity center and its interacting points. An interacting point is a portion of
molecule that could be affected by external forces: hydrophobia, electrostatic
charges, etc Fig.\ref{phospho} shows the generic 3D structure (part A) and its
application to a phospholipid design (part B). Forces and torques are applied
at the gravity center. Forces induce linear acceleration whereas torques induce
angular acceleration as in classical mechanics.\\

Spatial localization of each interacting point is defined by a vector in the
local frame associated with the agent (molecule) using its gravity center as
origin. This design gives information about the 3D structure of the molecule
and its space orientation but also its internal dynamics.

Modeling internal dynamics could be  important parameters in enzymatic
reactions. For example, an already filled active site will not be able to react
with another metabolite. A lot of enzymes interact with more than one
metabolite, in a certain order or randomly, before transforming them into new
products. The enzyme conformation could evolve after the binding of a first
substrate, revealing a second active site for a second type of substrate. Our
model allows these flexibles movements.

Even if their freedom is limited, the interacting points can be seen as agents
and thus molecules as communities of linked agents.

\begin{figure}[H]
    \begin{center}
        \includegraphics[scale=0.6]{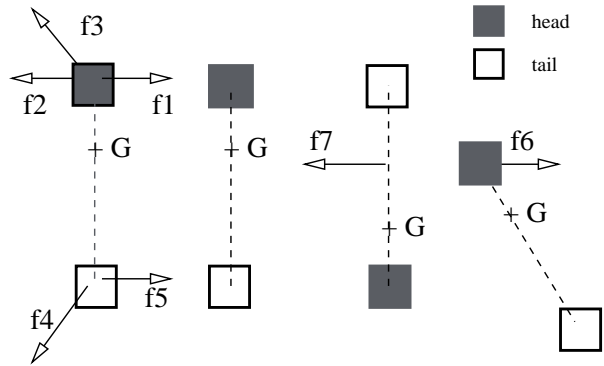}
        \caption{\label{forces}Example of interaction forces for
        phospholipids.}
    \end{center}
\end{figure}

Each pair of interacting points defines one direction with two types of
interaction - attraction/repulsion. Therefore, $n$ interacting points could
induce $2n(n-1)$ forces with intensity depending on the distance between
agents. As shown in Fig.\ref{forces}, the interactions between phospholipids
can be given by 4 forces $f_1, f_3, f_4, f_5$ that describe hydrophobia,
another one $f_2$ for hydrophilic and electrostatic charges, and the two last
ones $f_6, f_7$ for the molecular incompressibility.

\begin{figure}[H]
    \begin{center}
        \includegraphics[scale=0.6]{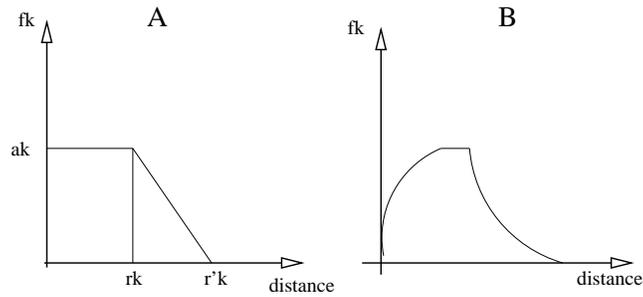}
        \caption{\label{forces-functions}Forces intensities are function of the
        distance. A: simple ramp function. B: example of other function available.}
    \end{center}
\end{figure}

As a first step, we have implemented simple functions with three parameters
(Fig.\ref{forces-functions} A): $a_k$ - maximal magnitude value of $f_k$, $r_k$
- the radius value under which  $f_k$ magnitude is maximal - and $r'_k$ - the
radius value over which $f_k$ is null. A second step will be to study the
impact of the choice of functions over phospholipid structure formation that
would emerge from MAS simulations. Fig.\ref{forces-functions} B shows another
function possibility.
\\

\newpage
With this approximation, the interactions between agents are reduced to a set
of forces. For a molecule $i$ in interaction with its neighbors, the compound
force is:
$$\overrightarrow{F}_i(t)=
\underbrace{a_{lin}\sum_h \sum_{j\neq i}\sum_k
\overrightarrow{f}_{h,k}(t)}_{\alpha_{lin}}
+\underbrace{b_{lin}\overrightarrow{rand}_{lin}(t)}_{\beta_{lin}}
-\underbrace{c_{lin}\overrightarrow{V}_i(t)}_{\gamma_{lin}}$$

\begin{list}{$\triangleright$}{where:}
\item $\alpha_{lin}$ is the forces from the neighborhood (forces from the $k$ interacting points
of the molecule $j$ acting upon the $h$ interacting points of the molecule
$i$),
\item $\beta_{lin}$ is the thermal effects depending on molecule type,
\item $\gamma_{lin}$ is the friction ($\overrightarrow{V_i}$ = molecule velocity) depending on molecule
type.
\end{list}
\paragraph{}
Linear and rotational accelerations are thus given by Newton's Second Law - the
fundamental law of dynamics: the acceleration of an object of constant mass is
proportional to the resultant force acting upon it (
$\overrightarrow{F}_i(t)=m\overrightarrow{a}$). This compound force is used in
the power series (order 2) of the position function $\overrightarrow{X}_i(t)$: 
%
%
%
\begin{eqnarray*}
    \overrightarrow{X}_i(t+\delta
t)&=&\overrightarrow{X}_i(t)+\frac{d\overrightarrow{X}_i(t)}{dt}\delta
t+\frac{d^2\overrightarrow{X}_i(t)}{2dt^2}\delta t^2+o(\delta t^2)\\
    \Leftrightarrow \Delta \overrightarrow{X}_i(t+\delta
t)&=&\overrightarrow{V}_i(t)\delta t+\frac{1}{2}\overrightarrow{a}_i(t)\delta
t^2+o(\delta
  t^2)\\
    \Leftrightarrow \Delta \overrightarrow{X}_i(t+\delta
t)&=&\overrightarrow{V}_i(t)\delta t+\frac{1}{2m}\overrightarrow{F}_i(t)\delta
t^2+o(\delta
    t^2)
\end{eqnarray*}
%
%
%
Each time step of the simulation requires for the linear movement:
\begin{displaymath}
    \left\{
        \begin{array}{rcl}
            \Delta \overrightarrow{V}_i(t+\delta t)&\approx &\frac{1}{m}\overrightarrow{F}_i(t)\delta t\\
            \Delta \overrightarrow{X}_i(t+\delta t)&\approx &\overrightarrow{V}_i(t)\delta t +\frac{1}{2m}\overrightarrow{F}_i(t)\delta t^2
        \end{array}
    \right.
\end{displaymath}
%

\begin{list}{$\triangleright$}{where:}
\item $m$ is the molecule mass,
\item $\overrightarrow{X}_i$ is the vector position of the gravity
  center of the molecule (i.e.: the origin of the local frame).
\end{list}

The forces applied by neighboring molecules induce torques. The compound torque
is:
$$\overrightarrow{T}_i(t)=
\underbrace{a_{rot}\sum_h \sum_{j\neq i}\sum_k
\overrightarrow{\tau}_{h,k}(t)}_{\alpha_{rot}}
+\underbrace{b_{rot}\overrightarrow{rand}_{rot}(t)}_{\beta_{rot}}
-\underbrace{c_{rot}\overrightarrow{W}_i(t)}_{\gamma_{rot}}$$

\newpage\noindent
\begin{list}{$\triangleright$}{where:}
\item $\alpha_{rot}$ is the torques from the neighborhood (torques from the $k$ interacting points
of the molecule $j$ acting upon the $h$ interacting points of the molecule
$i$),
\item $\beta_{rot}$ is the thermal effects depending on molecule type,
\item $\gamma_{rot}$ is the rotational friction depending on molecule type.
\end{list}

In the same way, each time step of the simulation requires for the angular
movement:
\begin{displaymath}
    \left\{
        \begin{array}{rcl}
            \Delta \overrightarrow{W}_i(t+\delta t)&\approx &\frac{1}{I}\overrightarrow{T}_i(t)\delta t\\
            \Delta \overrightarrow{\Theta}_i(t+\delta t)&\approx &\overrightarrow{W}_i(t)\delta
            t+\frac{1}{2I}\overrightarrow{T}_i(t)\delta t^2
        \end{array}
    \right.
\end{displaymath}
\begin{list}{$\triangleright$}{where:}
\item $I$ is the rotational inertia linked to the molecule size,
\item $\overrightarrow{\Theta}_i$ is the vector perpendicular to the rotation
plan, oriented so that the rotational movement is positive and which has a norm
corresponding to the angular variation.
\end{list}

This model with interaction points and gravity center seems to be a right way
to take into account the internal dynamics, the 3D structure and the space
orientation of the different biological molecules. Moreover the possibility to
adapt the agent granularity gives the possibility to design heterogeneous
molecules.

\section{Application to  the mitochondrial membranes and mitochondrial metabolism}

In a first attempt, we used this model to design mitochondrial membranes. Two
embedded membranes appear as an important characteristic of mitochondria. The
inner membrane is the more irregular one. As can be seen in Fig.\ref{mito} one
can understand that giving a mathematical definition of this form will be
rather difficult. On the other hand, using an emergent property of MAS
(analogous to natural membrane formation) can presumably be a better solution.
It can take into account the diversity of phospholipids constituting the
bilayer and the fact that the inner mitochondrial membrane is also composed of
a large amount of proteins complexes.

MAS also appear well suited to simulate the respiratory chain reactions,
particularly the electron transport, which occurs inside the inner membrane and
the proton transport which occurs across it. In principle, the mitochondrial
application of our MAS model should allow the natural emergence of the complex
electron transport by CoenzymeQ (CoQ) described under the name of
"Q-cycle"\cite{Mit76}, based on the reactions rules described for each CoQ site
on complexIII.

There is another field, in which MAS would certainly have a great interest ; it
is the field of mitochondrial diseases due to deficiencies in respiratory
complexes. Mitochondrial diseases are a group of complex pathologies, which can
affect different unrelated tissues with an unexpected association of symptoms
(\cite{Mau85}, \cite{Wal93}).

The disturbance in the distribution of the various \textsc{CoQ} species due to
mutations in respiratory chain complexes (giving sometimes only mild
respiratory chain deficiency) seems to be at the origin of (severe)
mitochondrial diseases \cite{Rus02}. Our MAS model easily describes the
distribution of the various redox CoQ species during the evolution of the
reactions, for any state of the respiratory complexes. It thus will give a
strong help to describe the plausible properties of the system and to design
new experiments allowing us to better understand these complex pathologies.

Finally reduced CoQ is an antioxidants \cite{Ern94}. Chemical derivatives of
this molecule are studied in order to reinforce its antioxidant properties and
to treat the oxidative stress in some mitochondrial diseases or during aging.
Our MAS model will help specifying the relevant targets of these molecules, to
keep them in their antioxidant form and thus help designing their properties.

\section*{Acknowledgment}
This work has been supported by the ACI IMPBio. The authors would like to
extend their thanks to Robert Strandh for its attentive relecture and its
judicious remarks.

\bibliographystyle{ieee}

\bibliography{main}
\end{document}